\documentstyle[prb,aps,epsf,psfig]{revtex}
\begin{document}
   \oddsidemargin -1 cm \topmargin -0.8 cm  
\twocolumn[\hsize\textwidth\columnwidth\hsize
           \csname @twocolumnfalse\endcsname
\title{Far infrared thermal spectroscopy 
of low-$T_c$ and high-$T_c$ superconductor films}
\author{R.~Tesa\v{r}, J.~Kol\'a\v{c}ek},
\address{Institute of Physics ASCR,
  Cukrovarnick\'a 10, 16253 Prague 6, Czech Republic}
\author{E. Kawate},
\address{National Institute of Advanced Industrial Science and 
    Technology 1-1-1, Central~2, Umezono, Tsukuba, Ibaraki 305-8568, Japan}
\author{\v{S}.~Be\v{n}a\v{c}ka, \v{S}.~Ga\v{z}i}
\address{Institute of Electrical Engineering SAS, Bratislava, Slovakia}
\maketitle
\begin{abstract}
Temperature dependence of far-infrared transmission of NbN thin films 
deposited on MgO and Si substrates was measured at several frequencies 
from 0.4 up to 4.3 THz. Exponential increase of relative penetration 
depth at low temperatures and a peak in transmission near $T_c$ was observed 
for frequencies below the optical gap. On the contrary, transmission 
measured at frequencies above the gap exhibits only flat, almost linear 
temperature dependence. This behavior is consistent with the BCS theory 
of superconductivity. Similar measurements were performed also on 
$\rm YBa_2Cu_3O_{7-\delta}$ 
thin films deposited on MgO and sapphire substrates. Low-temperature 
variation of transmission is linear or constant and the peak below $T_c$ 
predicted by the BCS theory is not observed. Flat temperature dependence 
at higher frequencies indicates that the photon energy was sufficient 
for excitation over the optical gap. The {\em s-wave} BCS theory is adequate 
for NbN films but not for the YBCO materials.
\end{abstract}
\vskip2pc]

\section{INTRODUCTION}
Far-infrared spectroscopy is an important tool for determining 
various parameters of superconductors. In the past it served as 
a direct evidence for an energy gap $2\Delta$ in the electronic excitation 
spectrum of the superconducting state \cite{57Glover}. The microscopic 
origin of the gap was clarified by the theory of Bardeen, Cooper and 
Schrieffer \cite{57BCS} which also successfully explained superconductor 
properties in the static electromagnetic field such as the Meissner 
effect. Subsequently, Mattis and Bardeen \cite{58MB} extended the theory 
to time-dependent electromagnetic interaction and obtained expressions 
for real and imaginary parts of the complex conductivity. 
Furthermore, other authors considered effects of impurities 
\cite{83Leplae,91Zimmermann} and strong coupling \cite{67Nam}. 
Soon after the discovery of high-temperature superconductivity it 
became obvious that the BCS-based theories supposing {\em s-wave} pairing 
of particles are not suitable for copper oxide materials. It is proposed 
that the order parameter $\Delta$ has $d$ or mixed $s$+$d$ symmetry 
\cite{98Modre}. The symmetry of the order parameter inevitably influences 
also the optical properties. Great attention was paid to analysis of 
the optical spectra of classical and high-$T_c$ superconductors in 
the IR and FIR regions \cite{89Timusk,92Tanner,Kawate}. 
The spectra are usually measured at several fixed temperatures using 
continuous source of radiation. The complementary method presented in 
this paper is known as {\em laser thermal spectroscopy} \cite{81Holah}. 
The FIR photon energies $\hbar\omega$  are fixed and the temperature of 
the sample is swept through the superconducting transition. 

The optical properties of a thin film on plane parallel substrate can be 
computed from the recurrence relations for reflection and transmission 
coefficients \cite{55Heavens}. If the substrate is wedged the expressions 
eliminating the interference fringes are more suitable 
\cite{71Gabriel}. For simplicity we give here a simple formula for 
transmission of the superconductor film \cite{57Glover} which neglects 
the interference in the substrate
\begin{equation}  
T_S=\left|1+{\sigma d Z_0 \over n+1}\right| ^{-2}.
\end{equation}
Here $\sigma$, $d$, $n$, $Z_0$, are conductivity and thickness of the film, 
index of refraction of the substrate and impedance of free space, respectively.
In the case when $|\sigma d Z_0/(n+1)|\gg 1$  the ratio of the 
transmission of the film in the superconducting state $T_S$ to the 
transmission $T_N$ in the normal 
state with conductivity $\sigma_N$ is reduced to
\begin{equation}
{T_S\over T_N}={\left| \sigma_N \right|^2 \over \left| \sigma \right|^2} .
\end{equation}

At low temperatures, where the real part of conductivity 
$\sigma_1(T)$ is much smaller than the imaginary part $\sigma_2(T)$, 
the relative transmission can be expressed in terms of the penetration 
depth $\lambda(T)$ \cite{96Vaulchier}
\begin{equation}
{T_S\over T_N}=\mu_0^2\omega^2\lambda^4(T)\left| \sigma_N \right|^2 , 
\end{equation}
which is often mentioned in discussion of microwave experiments. 
It was found that the low-temperature properties of the penetration 
depth can be used to distinguish between various types of symmetry of 
the order parameter $\Delta$ \cite{98Modre}. The {\em s-wave} BCS model in 
the local limit predicts exponential increase of the relative penetration 
depth with temperature \cite{91Turneaure}
\begin{equation}
{\lambda(T)-\lambda(0)\over\lambda(0)}=\sqrt{\pi\Delta\over 2kT}
{\rm exp}\left(-{\Delta\over kT}\right),\hspace{0.5cm} T<0.5 T_c
\end{equation}
whereas the {\em d-wave} symmetry is manifested by linear or quadratic 
temperature variation \cite{98Modre}.

The low temperature properties of classical superconductors were 
experimentally studied mainly at microwave and radio frequencies. 
The exponential temperature dependence of surface resistance or penetration 
depth was verified for NbN \cite{91Oates} and other materials 
\cite{91Turneaure,70Halbritter,97Trunin}. The FIR transmission of NbN 
films was fitted well using the Mattis-Bardeen theory \cite{82Perkowitz}.
The high-$T_c$ superconductors were investigated also in the far-infrared 
region but the results are still ambiguous. The temperature variation 
of the relative penetration depth strongly depends on the quality of 
sample; it was found to be constant, linear or 
quadratic \cite{96Vaulchier,96Wu,91Annet}. 

Near the critical temperature the Mattis-Bardeen theory predicts a 
sharp peak in transmission at frequencies comparable with the optical gap. 
Some indistinct maximum just below $T_c$ was already observed in classical 
superconductors by Perkowitz, who measured the temperature dependence of 
far infrared transmission of NbN films, one homogenous and one 
highly granular \cite{82Perkowitz}. The peak in transmission predicted 
by the theory was observed at a reduced level in the homogeneous sample, 
but it was absent in the granular material. In homogenous $\rm{V}_3Si$, 
which is a less strongly-coupled superconductor a higher maximum 
below $T_c$ was seen and better agreement with the same theory was achieved 
\cite{81Holah}. The presence of the peak in the copper oxide superconductors 
was not observed yet \cite{95Vaulchier}. It is the aim of this work to 
compare the temperature dependence of far-infrared transmission of low-$T_c$ 
and high-$T_c$ superconductor thin films. 

\section{THEORY}
In this section we will briefly describe some significant features of the 
applied theoretical model. The calculations are based on Zimmermann's 
expressions for complex conductivity of homogenous and isotropic BCS 
superconductor with arbitrary purity \cite{91Zimmermann}. 
Furthermore, the superconductor is supposed to be in the local limit 
($\xi_0\ll \lambda$), which is appropriate for both NbN and YBCO materials. 

\begin{figure}[h]
\label{fig1}
\centerline{\parbox[c]{11.3cm}{
\psfig{figure=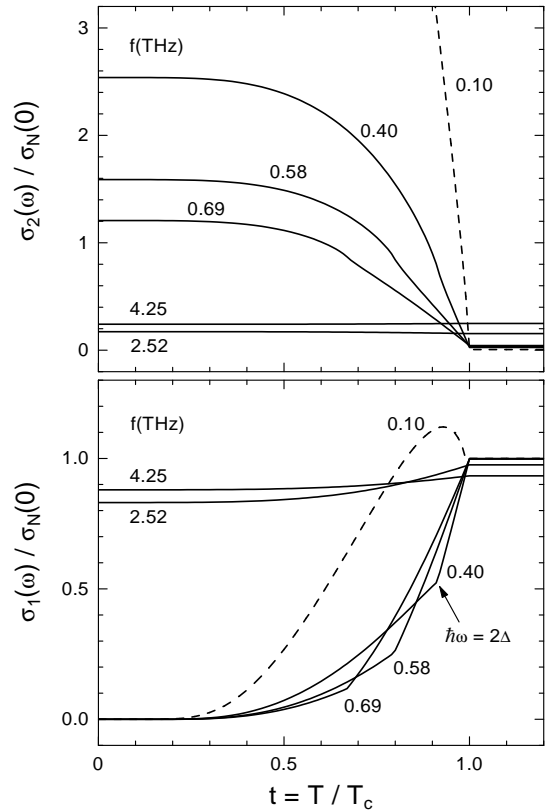,width=8cm,height=11.3cm}}}
\caption{
Theoretical temperature dependence of real and imaginary parts 
of the BCS complex conductivity. Parameters used in calculation: critical 
temperature $T_c$ = 10.8 K, dc conductivity in the normal state 
$\sigma_N(0) = 0.45 \;\Omega^{-1}\rm m^{-1}$, electron collision time 
$\tau = 10^{-14} \;\rm s$, optical gap $2\Delta_0 = 3.53 \; kT_c$.}
\end{figure}
The theoretical temperature dependence of the complex conductivity is 
displayed in fig. 1 for several frequencies in the range of 
interest and parameters relevant to NbN listed below the figure. 
Real and imaginary parts of conductivity are almost temperature 
independent for photon energies above the optical gap 
($\hbar\omega\gg 2\Delta_0$). For photon energy comparable with the 
gap ($\hbar\omega \approx 2\Delta_0$) the temperature behavior can be 
understood in the framework of the simple {\em two-fluid model}: Below $T_c$ 
the imaginary part increases with decreasing temperature and the real part 
declines to zero. At temperature $T_\omega$ for which photon energy 
coincides with the gap $\hbar\omega = 2\Delta(T_\omega)$ a discontinuity 
of slope appears on the $\sigma_1$ curves. At low enough frequencies 
($\hbar\omega \ll  2\Delta_0)$ a ``coherence peak'' arises in the real part 
of conductivity \cite{75Tinkham} (dashed line).

The complex conductivity $\sigma=\sigma_1+i\sigma_2$ presented in the 
previous paragraph is used to calculate the optical properties of 
a free standing thin film. From the relative complex permitivity 
$\epsilon_r=\epsilon_\infty+i\sigma/\omega$ we get 
the complex index of refraction which enters the Airy's formulas for 
reflection and transmission coefficients \cite{55Heavens}. Computed 
optical parameters are displayed in fig. 2.
\begin{figure}[h]
\label{fig2}
\centerline{\parbox[c]{11.3cm}{
\psfig{figure=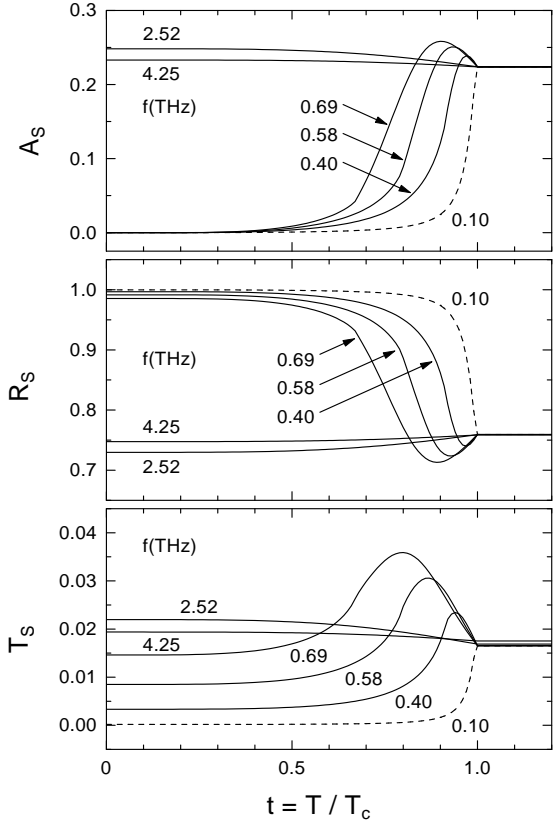,width=8cm,height=11.3cm}}}
\caption{ 
Optical properties of a free standing superconductor film: 
transmission $T_S$, reflection $R_S$ and absorption $A_S$. 
Material parameters same as in fig. 1, film thickness 
$d_{\rm film} = 80 \;\rm nm$, 
$\epsilon_\infty = 0$.} 
\end{figure}

In the high frequency limit, $\hbar\omega \gg 2\Delta_0$, the optical 
properties exhibit only flat, almost linear temperature dependence, not much 
different from the normal state. In the low frequency limit, 
$\hbar\omega \ll 2\Delta_0$, the optical parameters change from 
the normal state values with decreasing temperature. Transmission and 
absorption fall monotonously to zero whereas reflection comes to 100\% 
at zero temperature. At frequencies slightly lower then gap, 
$\hbar\omega \widetilde < 2\Delta_0$, a maximum appears in transmission and 
absorption just below $T_c$, while reflection exhibits a minimum. 
The peak in the optical properties 
does not correspond to the low frequency coherence peak in the real 
part of conductivity. Even monotonous dependence of the real and imaginary 
parts of conductivity may result in a peak in the optical properties if 
the real part of conductivity decreases more quickly than the imaginary part 
increases.

\section{EXPERIMENT}
The temperature dependence of FIR transmission was studied on two classical 
superconductor samples and two copper oxide samples. A high quality 
epitaxial NbN film (A) was deposited on MgO substrate and a polycrystalline 
NbN thin film (B) on Si substrate. A $\rm YBa_2Cu_3O_{7-\delta}$ thin film (C)
was prepared from a stoichiometric target by magnetron diode sputtering on 
MgO polished substrate oriented in the (100) plane. The c-axis of the grown 
film was directed perpendicularly to the sample plane. In order to eliminate 
light interference the substrate was slightly wedged. Another 
$\rm YBa_2Cu_3O_{7-\delta}$ film (D) was deposited on sapphire substrate 
covered by a very thin $\rm CeO_2$ buffer layer. The substrate was slightly 
wedged to suppress interference effect. Experimental parameters of the
measured samples are summarized in Table 1.  

The results presented in this paper were obtained mostly using the far 
infrared laser based spectrophotometer \cite{93Kolacek} in the Institute 
of Physics in Prague except for the measurements on the high quality NbN 
film (A) which were performed on a similar equipment \cite{Kawate} 
in the National Research Laboratory of Metrology in Tsukuba. We 
will briefly describe the former experimental set-up (for details see 
\cite{93Kolacek}). The source of the FIR radiation is an optically pumped 
gas laser which emits linearly polarised light at discrete frequencies 
in the range from 0.4 up to 4.3 THz (13-140 $\rm cm^{-1}$, 1.7-18 meV, 
742-70 $\mu \rm m$). Stability of the output laser power is monitored by a 
pyroelectric detector and the radiation transmitted through the sample 
is detected by a silicon bolometer kept at the working temperature of 
1.7 K. Transmission is proportional to the ratio of the signals from the 
bolometer and from the pyroelectric detector. As proved by careful 
testing, the sample holder is free of optical leakage. Background radiation 
is suppressed by an IR low pass filter inserted before the sample. 
Temperature of the sample can be controlled continuously in the range 
4.2 - 130 K and it is measured by a GaAs diode sensor attached to the 
sample holder. A small amount of helium exchange gas admitted into the 
sample space ensures good thermal contact. To enable correction of 
little difference between the sample and sensor temperatures the ac 
susceptibility at 330 Hz was monitored simultaneously with the transmission 
experiments. By this method the true superconducting phase transition 
is determined independently and the temperature axis rescaled accordingly.
\begin{table}
\label{tab1}
\caption{
Parameters $d_{\rm film}$, $T_c$, $\rho_N$ listed in the table 
represent the film thickness, the critical temperature, 
the dc resistivity in the normal state 
just above $T_c$. The substrates are in the shape of plate or wedge with 
dimensions $10 \times 10 \times d_{\rm sub} \;\rm mm^3$ where the thickness 
$d_{\rm sub}$ or the extent of the wedge is given in the last column.}
\begin{tabular}{|ccccccc|}
\hline
Sample & Film & $d_{\rm film}$ & $T_c$ &    
$\rho_N$ & Substrate/ & $d_{\rm sub}$ \\
&&[nm]&[K]&$[\mu\Omega {\rm cm}]$ & Buffer layer &[mm]\\
\hline 
A& NbN &    410 & 14.7&    71& MgO                 & 0.50\\ 
B& NbN &  70-90 & 10.8&   220& Si/$\rm SiO_2$      &  0.25\\
C& YBCO& 200-220&  89 &    ? & MgO                 & 0.37-0.98\\ 
D& YBCO& 120-140&  87 &   100& $\rm Al_2O_3/CeO_2$ & 0.36-0.67\\ 
\hline
\end{tabular}
\end{table}

\section{RESULTS}

The experimental results are presented in figures 3-7 together with the 
theoretical estimates. The temperature dependence of the relative 
transmission $T_S/T_N$ is plotted in the reduced temperature scale 
$t = T/T_c$. The intensity transmitted through the sample is normalized 
to that in the normal state taken at a temperature slightly above the 
superconducting transition. 
\begin{figure}[h]
\label{fig3}
\centerline{\parbox[c]{11.3cm}{
\psfig{figure=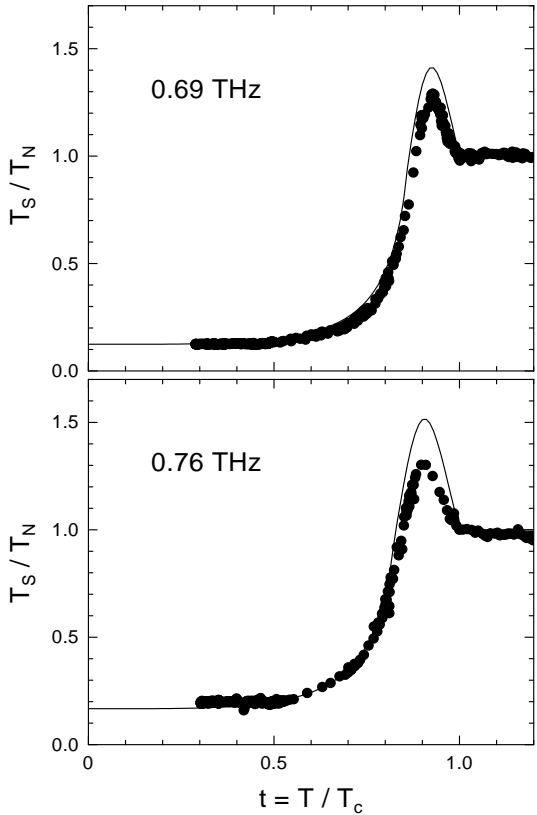,width=8cm,height=11.3cm}}}
\caption{
Temperature dependence of the relative transmission of NbN 
film A - experiment (points) and theory (lines). Parameters used in 
calculation: critical temperature $T_c = 14.7 \;\rm K$, dc conductivity in 
the normal state $\sigma_N(0) = 1.4\times 10^6\;\Omega^{-1}\;\rm m^{-1}$, 
electron collision time $\tau = 0.5\times 10^{-14} \;\rm s$, optical gap 
$2\Delta_0 = 3.53 \; kT_c$, film thickness $d_{\rm film} = 410 \;\rm nm$, 
substrate thickness $d_{S} = 0.5 \;\rm mm$, 
index of refraction of the substrate 
$n_{\rm sub} = 4.85 + 1.5 \times 10^{-7}i$.}
\end{figure}

The measurements performed on the high 
quality NbN film (A) at laser frequencies 0.69 and 0.76 THz are shown in 
fig. 3. The transmission peak just below $T_c$ is sharp and well 
pronounced. At very low temperatures the shape of the dependence was not 
reliably resolved, as due to the relatively high film thickness the 
transmitted power was small. 
\begin{figure}[h]
\label{fig4}
\centerline{\parbox[c]{11.3cm}{
\psfig{figure=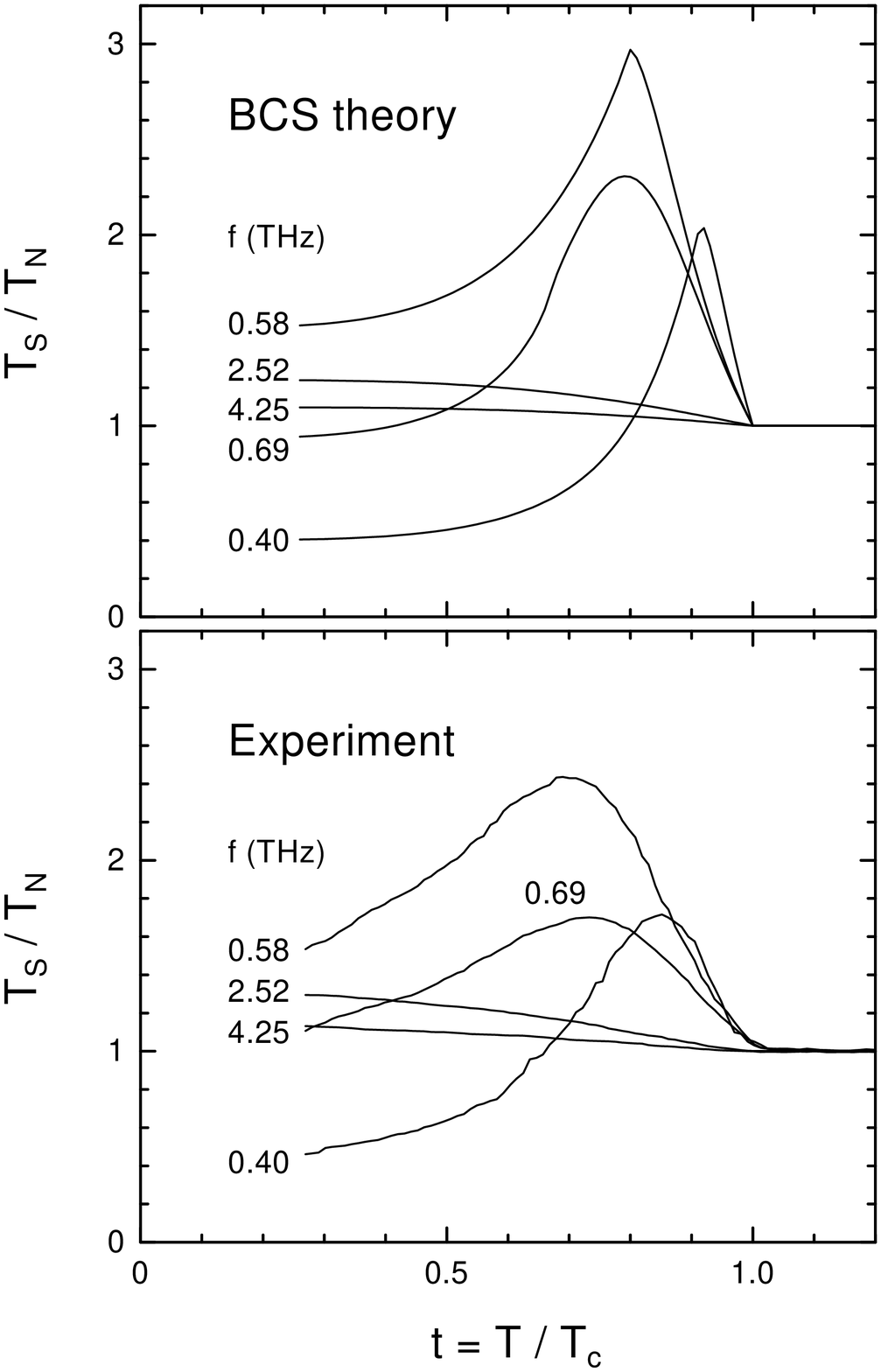,width=8cm,height=11.3cm}}}
\caption{
Temperature dependence of the relative transmission of NbN thin film 
B - experimental and theoretical. Parameters used in calculation: critical 
temperature $T_c = 10.8 \;\rm K$, dc conductivity in the normal state 
$\sigma_N(0) = 0.45\times 10^6 \;\Omega^{-1}\rm m^{-1}$, electron collision 
time $\tau = 10^{-14}\;\rm s$, optical gap $2\Delta_0 = 3.53 \; kTc$, 
film thickness $d_{\rm film} = 80 \;\rm nm$, substrate 
thickness $d_{\rm sub} = 0.25\;\rm mm$, index of refraction of the substrate 
$n_{\rm sub} = 3.5$.}
\end{figure}

The polycrystalline NbN thin film (B) exhibits a broad peak near $T_c$ at 
laser frequencies 0.4, 0.58 and 0.69 THz but a flat dependence at 
frequencies 2.52 and 4.25 THz (see fig. 4). The relative penetration 
depth displayed in fig. 5 was calculated from the low temperature 
measurement at frequency 0.4 THz. The experimental data were fitted 
using the exponential expression (4) which follows from the BCS theory.
\begin{figure}[h]
\label{fig5}
\centerline{\parbox[c]{6.45cm}{
\psfig{figure=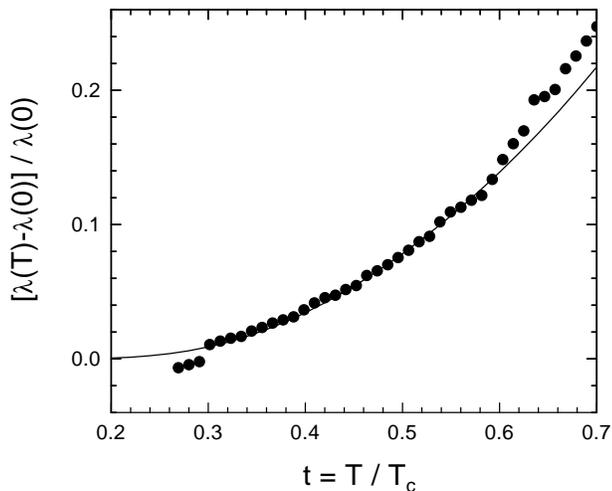,width=8cm,height=6.45cm}}}
\caption{
Temperature dependence of the relative penetration depth of NbN 
thin film B - experiment (points) and the best fit (line) with the gap 
value $2\Delta_0 = 3.53 \;kT_c$.}
\end{figure}

\begin{figure}[h]
\label{fig6}
\centerline{\parbox[c]{11.3cm}{
\psfig{figure=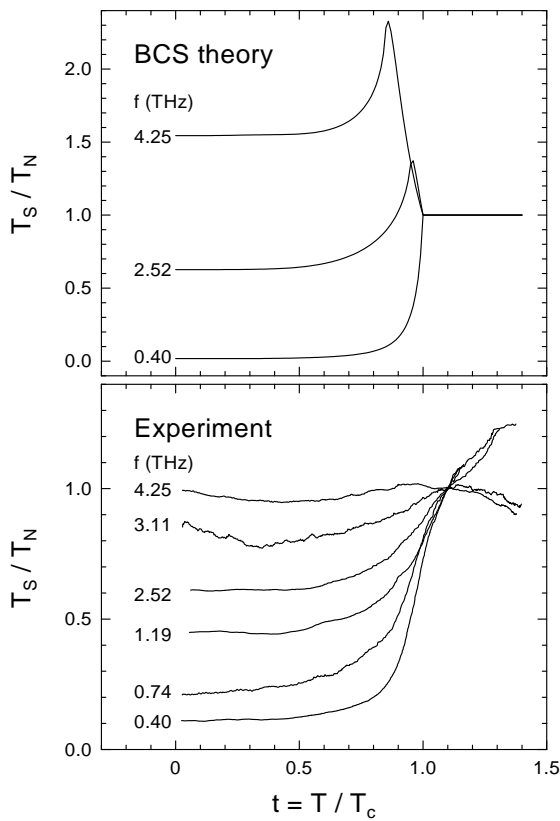,width=8cm,height=11.3cm}}}
\caption{
Temperature dependence of the relative transmission of 
$\rm YBa_2Cu_3O_{7-\delta}$ thin film C - measured and calculated from the 
BCS theory with parameters: $T_c = 89\; K$, dc conductivity in 
the normal state $\sigma_N(0) = 10^6 \;\Omega^{-1}\rm m^{-1}$, 
electron collision time $\tau = 3.5 \times 10^{-14}\;\rm s$, 
optical gap $2\Delta_0 = 3.53 \;kT_c$, film thickness 
$d_{\rm film} = 210\;\rm nm$, 
substrate thickness $d_{\rm sub} = 0.5 \;\rm mm$, 
index of refraction of the substrate 
$n_{\rm sub} = 4.85 + i \;1.5\times 10^{-7}$.}
\end{figure}

At low temperatures the transmission of the YBCO sample (C) changes 
slightly or is almost constant (see fig. 6). At the vicinity of 
$T_c$ a clear step appears in transmission for lower frequencies 0.40-2.5 THz.
The magnitude of the step diminishes towards higher frequencies 
and above 3.11 THz the dependence on temperature becomes nearly flat. 
The sharp peak expected by the BCS theory is absent.
Similarly for the YBCO film (D) the transmission grows monotonously up to 
$T_c$ without any maximum (see fig. 7). 
\begin{figure}[h]
\label{fig7}
\centerline{\parbox[c]{6.45cm}{
\psfig{figure=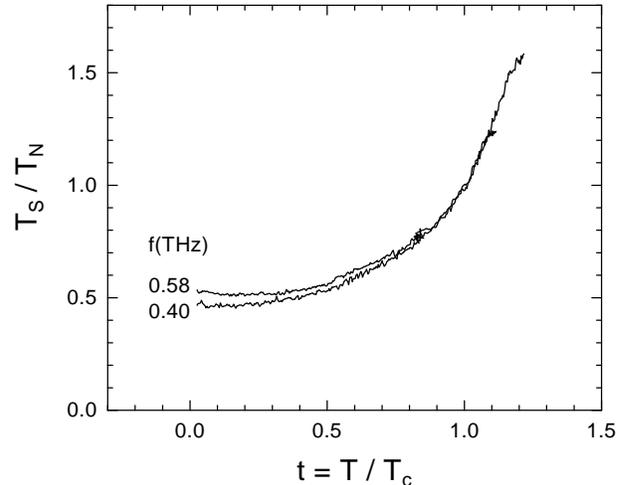,width=8cm,height=6.45cm}}}
\caption{
Temperature dependence of the relative transmission of 
$\rm YBa_2Cu_3O_{7-\delta}$ thin film D.}
\end{figure}

The theoretical curves are calculated with parameters listed in the 
figure captions. In particular the experimental data from table 1 are 
exploited, the gap is fixed at the BCS value $2\Delta_0=3.53 \;kT_c$, 
for electron collision time $\tau$ a reasonable guess is used.  
High frequency permitivity $\epsilon_\infty$ is neglected
and the index of refraction of the substrate $n_{\rm sub}$ is taken from the 
literature.
In the theoretical estimates of the FIR transmission of the NbN samples (A,B) 
also the multiple reflections in the substrate are included 
\cite{55Heavens}. Interference of the FIR radiation affected 
qualitatively only the transmission of the NbN sample B (fig. 4), 
for which an irregularity in the curve ordering should be noticed.
For both YBCO samples (C,D) which are deposited on wedged substrates 
the formula neglecting intererence effect was used \cite{71Gabriel}.

\section{CONCLUSIONS}
In the theoretical model the following feature was observed for the range 
of applied parameters: if the real part of conductivity exhibits 
a maximum below $T_c$ the optical parameters change monotonously with 
temperature and vice versa.

The temperature dependence of the far-infrared relative transmission of 
classical and high-temperature superconductor films is substantially different.
The NbN experimental data are well described by calculations based on the 
{\em s-wave} BCS model. In the NbN samples, the exponential increase of 
relative penetration depth at low temperatures and the peak in 
transmission near $T_c$ at frequencies below the optical gap were observed. 
The peak strongly depends on the quality of the sample. In the high quality 
sample (A) it corresponds well to BCS calculations, whereas in the 
polycrystalline sample (B) the observed maximum is somewhat broader and 
its level a little lower than theoretical. This behavior is in accordance 
with other experiments which were performed on granular films 
\cite{82Perkowitz}. In the YBCO samples, the low temperature variation of 
transmission is non-measurably small and the peak below $T_c$ predicted by 
the BCS theory is missing. The flat temperature dependence at laser 
frequency 4.3 THz (fig. 6) indicates that the photon energy was 
sufficient for excitation over the optical gap. The {\em s-wave} BCS theory 
is adequate for NbN films but it is not appropriate for the YBCO materials. 

{\bf Acknowledgments.} 
The authors are grateful to V. Gregor and \v{S}. Chromik 
for preparing the YBCO samples and to V. Kambersk\'y and J. Loos for useful 
discussion. This work was supported by the Grant Agency of the ASCR 
under contract \#A1010919 and by the Slovak Grant Agency under 
contract \#2/7199/20. The European ESF program VORTEX is also gratefully
acknowledged.


\end{document}